\documentclass[11pt]{article}

\usepackage{cite}
\usepackage{amssymb}
\usepackage{amsmath}
\usepackage{bm}

\begin{document}

\title{Quantum Processes and Energy-Momentum Flow.}
\author{B. J. Hiley\footnote{E-mail address b.hiley@bbk.ac.uk.}  and D. Robson.}
\date{TPRU, Birkbeck, University of London, Malet Street, \\London WC1E 7HX.\\ \vspace{0.4cm}(20 November 2014) }
\maketitle

\begin{abstract}
	In this paper we focus on energy flows in simple quantum systems.  This is achieved by concentrating on the quantum Hamilton-Jacobi equation.  We show how this equation appears in the standard quantum formalism in essentially three different but related ways, from the standard Schr\"{o}dingier equation, from Lagrangian field theory and from the von Neumann-Moyal algebra.  This equation allows us to track the energy flow using the energy-momentum tensor, the components of which are related to weak values of the four-momentum operator.  This opens up a new way to explore these components empirically.  The algebraic approach enables us to discuss the physical significance of the underlying non-commutative symplectic geometry, raising questions as to the structure of particles in quantum systems.

\end{abstract}

\section{Introduction}

The experimental results using weak measurements performed on single photons by Kocsis {et al} \cite{kas11} in which they claim to construct ``photon trajectories" re-opens the debate concerning the relation between the individual and its field.  The classic discussions between Bohr and Einstein reported so clearly in Jammer~\cite{mj74} have never been finally resolved; rather there has been a tendency to discourage re-opening what appears to be an unending tension between the various views.  Now with new information both theoretical and experimental it is time to re-evaluate the position.

Einstein's question as to whether one should work with the wave function $\psi$ alone or add some detailed specification of the localisation of the particle remains unanswered; each adopting their own private preferences at treating individual questions that arise when using the quantum algorithm.  Fortunately  liberal use of the uncertainty principle can suppress attempts to go `deeper' but the appearance of weak values has raised these very real issues again~\cite{bbkn13, fh12}.  My own researches~\cite{dbbh93} have tended to involve exploring the consequences of adding a localisation feature, not because I believed it was indispensable, Bohr has already shown that is not the case, but more because I could not understand why there was such an antipathy against exploring the question.  The norm was `Use the rules and you find answers to the questions the rules allow and Nature endorses!'  With the advent of measurable weak values we can ask new questions which can be put to the experimental test, and it is these questions that I want discuss in this paper.

\section{The usual Formalism but where is the Energy?}

We start with a quantum system described by a wave function, $\psi(x,t)$, the time evolution of which is given by the Schr\"{o}dinger equation. The Born probability rule is then used to calculate the probability $P(x',t')$ of finding the system at $x'$ at a later time $t'$.  Thus $P(x',t') = |\psi(x',t')|^2=R^2(x',t)$ where $R(x',t)$ is the amplitude of the field.
 So our final result  depends only on one of the pair of real numbers in $\psi(x,t)=R(x,t)e^{iS(x,t)/\hbar}$.    The information as to how the phase evolves in time is, as it were,  `hidden' in the evolution of the complex wave function $\psi(x,t)$.  It would perhaps be revealing to have a pair of equations showing explicitly the evolution of the two real fields $R(x,t)$ and $S(x,t)$.

The simplest way to arrive at the equations containing $R$ and $S$ is to substitute $\psi=Re^{iS/\hbar}$ into the Schr\"{o}dinger equation and separate the resulting equation into
its real and imaginary parts.
The imaginary part can be written in the form\begin{eqnarray}
\frac{\partial\rho}{\partial t}+\nabla.\left(\rho\frac{\nabla S}{m}\right)=0.	\label{eq:Liouv}
\end{eqnarray}
 Since, at this stage, we are simply analysing the Schr\"{o}dinger equation, equation (\ref{eq:Liouv}) provides an expression for the conservation of probability $P(x,t)$. 
 The real part  takes the form\footnote{I am well aware I am here treating $S$ as a torsor.  For a discussion of the physical meaning of $S$ see p. 95 of Mackey~\cite{gm63}.}
\begin{eqnarray}
\frac{\partial S}{\partial t}+\frac{(\nabla S)^2}{2m}-\frac{\hbar^2}{2m}\left(\frac{\nabla^2R}{R}\right)+V=0.		\label{eq:qhj}
\end{eqnarray}
We will call this the quantum Hamilton-Jacobi equation [QHJ] for reasons that we will bring out as we go along.

These two equations must have the same content as the Schr\"{o}dinger equation and it would surely be of interest to see if they can give a different insight into the evolution of quantum systems.  Note we are not departing from the usual interpretation yet, we are merely drawing attention to an alternative form of the mathematical structure.  Already one sees that there is a disadvantage of using these two equations as they are no longer linear and therefore more difficult to analyse.  Nevertheless as we will show, we can obtain new information about energy flow using equation (\ref{eq:qhj}), in spite of Bohr's insistence that you can talk either about an evolution in space-time or about a causal (i.e. momentum-energy) evolution, never both together. 

Although the splitting of an equation into its real and imaginary parts is a standard mathematical practice,  we will re-derive these two equations again, starting from Heisenberg's expression for the Lagrangian of the \newline Schr\"{o}dinger field~\cite{wh49} and applying the standard Euler-Lagrange equations, treating $R(x,t)$ and $S(x,t)$ as independent fields.  This procedure  will enable us to find the components of the energy-momentum tensor, thus  allowing us to investigate the energy and momentum flows involved in the quantum process.  In this way  we are able show that equation (\ref{eq:qhj}) is an expression for the local conservation of energy in this evolving quantum process.

This result should not be too surprising since, as is well known, the Schr\"{o}dinger equation must describe  the evolution of the energy involved in the process.  Why?   Because the expression of the classical dynamical energy, the Hamiltonian, albeit written in operator form, is at the heart of the equation.  However by focussing on the complex form of the wave function, we do not explicitly see how this energy flows in the evolving process.  The wave function then appears, as it were, `disembodied'  from the energy, so that it then seems to take on, physically, the air of some ghostly shadow of the evolving system, allowing only probability outcomes to be discussed.  

We then find that the wave function, with its deterministic equation, can be treated as
 an entity in its own right  giving the probability of finding a particular result.  Its role in accounting for the energy flow is then forgotten.  In consequence we feel free to add wave functions and to collapse wave functions with no concern as to the energy involved, hoping that it will be taken care of by the Schr\"{o}dinger equation.  
 
 However a realisation that both the addition of wave functions and the collapse of wave functions occur outside of the Schr\"{o}dinger equation,  should be a cause for concern since, unless care is taken with such addition and collapse, any such move could contradict the conservation of energy\footnote{We are talking about energy non-conservation outside the limits imposed by the energy-time uncertainty principle.}.
The purpose of this paper is find a way to discuss the flow of energy in a quantum process rather than relying {\em only} on the $\psi(x,t)$ and the Schr\"{o}dinger equation.

\subsection{Energy of the Schr\"{o}dinger Field}

We will follow Heisenberg \cite{wh49} and write the Lagrangian, ${\cal L}(\psi,\partial_{\mu}\psi)$ for the Schr\"{o}dinger particle as
 \begin{eqnarray}
{\cal L}= - \frac{1}{2m}\nabla \psi^{*}\cdot\nabla \psi+\frac{i}{2}[(\partial_{t}\psi)\psi^{*}-(\partial_{t} \psi^{*})\psi]-V\psi^{*}\psi.	\label{eq:Lagrangian}
 \end{eqnarray}
 Let us remind ourselves of how this Lagrangian leads to the Schr\"{o}dinger equation.  All we have to do is to use the Euler-Lagrange equations, treating $\psi$ and $\psi^*$ as two independent fields,
 \begin{eqnarray*}
 \frac{\partial {\cal L}}{\partial \psi}-\partial_\mu\left(\frac{\partial {\cal L}}{\partial \psi_\mu}\right)=0\quad\mbox{and}\quad \frac{\partial {\cal L}}{\partial \psi^*}-\partial_\mu\left(\frac{\partial {\cal L}}{\partial \psi_\mu^*}\right)=0
\end{eqnarray*}
where $\psi_\mu=\partial_\mu\psi$ and $\psi^*_\mu=\partial_\mu\psi^*$.
If we use these relations together with
\begin{eqnarray*}
\frac{\partial {\cal{L}}}{\partial(\partial^{0}\psi)}=\frac{1}{2i}\psi^{*}
\hspace{0.5cm}\mbox{and}\hspace{0.5cm}
\frac{\partial {\cal{L}}}{\partial(\partial^{0}\psi^{*})}=-\frac{1}{2i}\psi,
\end{eqnarray*}
 we obtain {\em both} the Schr\"{o}dinger equation and its dual, the complex conjugate equation, viz,  
 \begin{eqnarray}
 i\hbar\frac{\partial\psi}{\partial t}= -\frac{\hbar^2}{2m}\nabla^2\psi +V\psi\quad\mbox{and}\quad  i\hbar\frac{\partial\psi^*}{\partial t}= \frac{\hbar^2}{2m}\nabla^2\psi^* -V\psi^*.		\label{eq:ss}
 \end{eqnarray}
 Let me stress, we obtain {\em two} equations using the Euler-Lagrange approach because we have treated $\psi$ and $\psi^*$ as independent variables.  This dual equation seems to add nothing physically new as it is simply the complex conjugate of the first equation and therefore contains no new information.

\subsection{Energy-momentum Tensor}

The energy-momentum tensor of any field with Lagrangian ${\cal L}$ is defined by
\begin{eqnarray}
T^{\mu\nu}=- \left\{\frac{\partial {\cal{L}}}{\partial(\partial^{\mu}\psi)}\partial^{\nu}\psi+\frac{\partial {\cal{L}}}{\partial(\partial^{\mu}\psi^{*})}\partial^{\nu}\psi^{*}\right\}-{\cal L}\delta^{\mu\nu}.	\label{eq:EM}
\end{eqnarray}
Using this expression, we find the momentum density of the Schr\"{o}dinger field can be written as
\begin{eqnarray*}
T^{0j}=-\left\{\frac{\partial {\cal{L}}}{\partial(\partial^{0}\psi)}\partial^{j}\psi+\frac{\partial {\cal{L}}}{\partial(\partial^{0}\psi^{*})}\partial^{j}\psi^{*}\right\},
\end{eqnarray*}
which reduces to
\begin{eqnarray*}
T^{0j}=\frac{i}{2}\left[\psi^{*}\partial^{j}\psi - \psi\partial^{j}\psi^{*}\right].
\end{eqnarray*}
We immediately recognise that $T^{0j}$ is a current for the momentum flux.  In terms of the real field $S(x,t)$, this momentum current can be written in the form
\begin{eqnarray}
T^{0j}=-\rho\partial^jS:={\cal G}^j.      \label{eq:BohmM}
\end{eqnarray}
It  should be noted that the momentum derived from  $T^{0\mu}$ is {\em not} the momentum given by one of the eigenvalues of the momentum operator when the wave function is $\psi(x,t)$.  It is information contained in the phase of the wave function and, being an energy, should be experimentally  accessible without violating the uncertainty principle, which applies only to the   simultaneous measurement of the eigenvalues of operators.
Indeed we will show that it is accessible through weak measurement and is the real part of a weak value of the momentum operator~\cite{rl05, bbkn13}.

This opens up a new experimental way to explore quantum processes, allowing us access to new physical information.  Not only that, but from the precise relationship $\rho P_B^j=T^{0j}$,  a weak measurement will allow us to measure $P_B=\nabla S$, a relation first introduced by de Broglie\footnote{Mackey~\cite{gm63} notes that ``$\partial S/\partial x$ describes the momentum of the state" of the particle.  Bliokh {\em et al.}~\cite{bbkn13} call it the local momentum.} and one that forms a key component of the Bohm approach as defined in Bohm and Hiley \cite{dbbh93}.  Thus the approach can be opened up  to experimental investigation\footnote {Let us note in passing that the momentum, $P_B=\nabla S$, introduced by Bohm~\cite{db52} and now used in Bohmian mechanics,  is simply related to the $T^{0j}$ component of the energy-momentum tensor.  The corresponding energy current density is obtained from the tensor component $T^{j0}$ (For details see Holland~\cite{ph95}.)  In Hiley~\cite{bh08} we have shown that the Bohm momentum is identical to the conditional expectation value of the momentum constructed from the Moyal probability distribution. Thus there is a much closer relation between these approaches than is normally realised.}~\cite{rfbh13}.   Note also that this approach does not contradict standard quantum mechanics, it is simply an exploration of a different aspect of it, showing that there is {\em more experimental content} available in addition to the eigenvalues found in von Neumann type measurements.

It is very important to distinguish between the Bohm momentum $P_B$ obtained from a weak measurement and the momentum eigenvalue $P_\psi$ that results from a strong measurement.  How is it possible for a particle to have two different values of momentum?  The weak measurement involves inducing a phase change in the spin state of the system~\cite{rfbh13}.  This change is found by performing a von Neumann measurement on the spin.  All strong measurements are participatory and do not passively reveal what is already there, unless the system is in an eigenstate of the operator involved in the measurement. Being participatory, they are not faithful and do not merely reveal what is already there.

In passing it should be noted that there are also connections with other components of the energy-momentum tensor.  Then we find
\begin{eqnarray*}
T^\mu_\mu=-\frac{\hbar^2}{m}[\nabla^k\psi^*\nabla_k\psi]-\frac{i}{2}[\psi^{*}\partial_{0}\psi-\psi\partial^{0}\psi^{*}].
\end{eqnarray*}
If we examine the last two terms in this trace we find
\begin{eqnarray*}
T^{00}+{\cal L}\delta^{00}=\frac{i}{2}[\psi^{*}\partial^{0}\psi-\psi\partial^{0}\psi^{*}]=-\rho\partial_tS=\rho E_B,
\end{eqnarray*}
which gives the  energy, $E_B=-\partial_tS$, that Bohm~\cite{db52} introduced in his approach.  Thus we see that the Bohm momentum and the Bohm energy are intimately related to the energy of the Schr\"{o}dinger field.
But there is more.  We also find
\begin{eqnarray*}
T^{kk}-{\cal L}\delta^{kk}=\frac{R^2}{2m}\left[(\partial^{x_k} S)^2+\hbar^2\frac{(\partial^{x_k}R)^2}{R^2}\right]+VR^2.
\end{eqnarray*}
Thus not only does the kinetic energy,  $(\nabla S)^2/2m $, emerge, but there also appears a new form of kinetic energy, namely, $KE_0=\hbar^2(\nabla R)^2/2mR^2 $.  This is the kinetic energy associated with what Nelson~\cite{en66} calls the `osmotic velocity', $v_0=\hbar\nabla R/mR$.

The appearance of this osmotic velocity has always been a curious feature of the formalism but the question as to what it means is generally ignored.  However we can now identify it with the imaginary part of the weak value of the momentum operator as we will show later. The important point to notice is that it can now be investigated experimentally using the techniques of weak measurement.

\subsection{Energy and Momentum Flows}

In order to determine the time development of the two real fields, let us now  write the Lagrangian (\ref{eq:Lagrangian}) in terms of $R(x,t)$ and $S(x,t)$.  We then find
\begin{eqnarray}
{\cal L}=-R\left(\frac{\partial S}{\partial t}+\frac{(\nabla S)^2}{2m}+\frac{\hbar^2}{2m}\frac{(\nabla R)^2}{R^2} +V\right).  	\label{eq:realLagrangian}
\end{eqnarray}
Again using the Euler-Lagrange equations with  $R$ and $S$ as independent fields, we find two equations, the first is the conservation of the probability equation (\ref{eq:Liouv})
\begin{eqnarray*}
\frac{\partial\rho}{\partial t}+\nabla.\left(\rho\frac{\nabla S}{m}\right)=0.	
\end{eqnarray*}
  The second equation gives equation (\ref{eq:qhj})
\begin{eqnarray*}
\frac{\partial S}{\partial t}+\frac{(\nabla S)^2}{2m}-\frac{\hbar^2}{2m}\left(\frac{\nabla^2R}{R}\right)+V=0.\hspace{2cm}(2)\hspace{-4.5cm
}	
\end{eqnarray*}
Not surprisingly these are exactly the two equations we obtained earlier simply substituting $\psi=Re^{iS/\hbar}$ into the Schr\"{o}dinger equation and separating the resulting equation into its real and imaginary parts. 

Our discussion of the energy-momentum tensor now allows us to identify equation (\ref{eq:qhj}) as an expression for the local conservation of energy.    Using the relation $\partial_t S(x,t)=-E_B(x,t)$ as the energy of the particle, and identifying $(\nabla S(x,t))^2/2m=(P_{B}(x,t))^2/2m$ as the kinetic energy of the particle, equation (\ref{eq:qhj}) takes the form
\begin{eqnarray*}
E_B= KE + QPE +CPE.
\end{eqnarray*}
In a closed system this equation is an expression for the {\em local} conservation of energy, provided we regard the quantum potential energy, $QPE=-\hbar^2\nabla^2R(x,t)/2mR(x,t)$, as a new {\em quality} of energy present only in the quantum domain.

At face value it appears that this equation calls into question Bohr's notion of complementarity--it is both causal and evolves in space-time.  However this is achieved by the appearance in the mathematics of the new quality of energy, namely, the QPE.  Clearly something very different is happening here, something that needs further investigation which we will take up in detail in section 5.  Before we go into the meaning of this new form of energy, let us first note how this equation fits in with classical mechanics and also with quantum field theory.

If the $QPE$ is taken to be zero, then equation (\ref{eq:qhj}) 
has a  similar form to the classical Hamilton-Jacobi equation
\begin{eqnarray}
\frac{\partial S_a}{\partial t}+\frac{(\nabla S_a)^2}{2m}+V=0. \label{eq:chj}
\end{eqnarray}
Here $S_a$ is the classical action.  This suggests that when we can neglect the term in $\hbar^2$, i.e., when the quantum potential energy (QPE) becomes negligible, and if we allow the phase $S$ to become the classical action $S_a$, we arrive at the  equation of motion of a classical particle.  This means that classical physics actually emerges from quantum physics in a very clear way.  This is of course, just an example of quantum deformation theory~\cite{ahph02}.

Another connection with the classical formalism emerges when we explore the Lagrangian approach further.   Notice that the momenta conjugate to $S$ and $\rho$ are $\pi_S=-\rho$ and $\pi_\rho=0$,  which means that the Hamiltonian density can be written in the form
\begin{eqnarray*}
{\cal H}=-\rho\partial_tS-{\cal L}=\rho\left[\frac{(\nabla S)^2}{2m}+\frac{\hbar^2}{8m}\frac{(\nabla \rho)^2}{\rho^2}+V\right].
\end{eqnarray*}
Regarding this Hamiltonian as a functional of $S$ and $\pi_S=-\rho$ and using Hamilton's equations of motion, we again obtain equations (\ref{eq:Liouv}) and (\ref{eq:qhj}).

The similarity in form of these two equations led Bohm~\cite{db52} to initially propose that we could retain the notion of a classical particle even in the quantum domain provided, as he argued, one adopts the classical canonical relations, $p=\nabla S_a$ and $E=-\partial_tS_a$ and simply replace $S_a$ by the phase $S$, a step that he did not justify.  He regarded them as `subsidiary conditions'~\cite{db53}.  This led him to propose initially that we could regard equation (\ref{eq:qhj}) as describing a point particle following a well defined trajectory. However in Bohm and Hiley~\cite{dbbh93}, we already noted that this approach implied that the particle could not be a classical point particle, but that it had some internal structure, a feature that we will explore further in this paper.

  \subsection{Relation to Quantum Field Theory}
  
  Before going on to discuss the significance of the appearance of the new term QPE in equation (\ref{eq:qhj}), it is necessary to appreciate the relation of the energy-momentum results we have obtained to those used in standard quantum field theory.
The energy-momentum tensor~(\ref{eq:EM}) is in fact  used in conventional quantum field theory (QFT) not only for the Schr\"{o}dinger field but also for the Dirac and Klein-Gordon fields~\cite{ss64}.  In this paper we will only discuss the Schr\"{o}dinger field, but our conclusions hold in general. 

Conventional QFT uses the same energy-momentum tensor to determine the energy and momentum of a particle, not in the way we have, but by forming the integral
\begin{eqnarray}
P^{\mu}=\int T^{0\mu}(x)d^{3}x.	\label{eq:momentum}
\end{eqnarray}
In other words conventional QFT identifies the energy and momentum of the particle with the {\em global} energy and momentum. 
The conservation of energy and momentum is then ensured  because
\begin{eqnarray}
\frac{d}{dx^{0}}\int T^{0\mu}(x)d^{3}x=0\quad\forall \mu.	\label{eq:conse}
\end{eqnarray}
Thus conventional quantum field theory works with the {\em global} properties of the field.  In contrast in this paper we identify the energy and momentum with the {\em local} properties of the field.  While equation (\ref{eq:conse}) is identified with the conservation of global energy-momentum, equation (\ref{eq:qhj}) is identified with the local energy-momentum conservation.  So once again we are exploring different aspects of exactly the same structure. 
  
  \subsection{ Energy Conservation.}
  
It was not until the '70s that a programme was started to systematically explore the consequences of the QHJ if we assume  that we can retain the notion of a localised particle and calculate trajectories in the same spirit as used for classical particles.  The key assumption here was the particle had simultaneously both a well defined position and momentum even though both could not be measured simultaneously.  The results of this exploration can be found in Bohm and Hiley~\cite{dbbh93} and Holland~\cite{ph95} so it is not necessary to the details of this work here.

Although we retained the notion of a local particle following a trajectory in Bohm and Hiley~\cite{dbbh93}, we made it clear that this did not mean we were retaining {\em all} the classical concepts.  The source of the novelty emerges from the appearance of the QPE.  This energy is of a totally different quality from a traditional classical potential energy.  It had no external source and depended on the experimental conditions in an irreducible way, in the sense that it is not a preassigned function of position. It behaved more like an informational potential for the reasons we discussed in detail in Bohm and Hiley~\cite{dbbh93A} and will not repeat here.  The way we envisioned it working would lead us to propose the quantum particle was not a point object, but would have internal structure that is very different from a classical point-like object.    

The fact that we were not returning to classical concepts had been pointed out by Bohm himself much earlier in chapter 5 of his book ``Causality and Chance" \cite{db57} where he already began to present a radically different view of quantum processes, a view that would hopefully take us further in  finding a clearer understanding of quantum phenomena that would open up the possibility of a quantum theory of gravity.  In the last chapter of Bohm and Hiley~\cite{dbbh93} we sketched briefly some development of these ideas and it was our intention to develop them into a second book but unfortunately Bohm died before we could complete the project.

It is also unfortunate that those who came later and developed what they call ``Bohmian mechanics"~\cite{ddsg96}  totally ignored Bohm's own views on this subject, claiming that one can maintain a mechanical viewpoint contrary to the fact that Bohm had eloquently argued against a mechanistic philosophy and had called for, and was actually exploring,  a more organic view of quantum phenomena~\cite{db82a}.

The results of section 2.1 clearly show unambiguously that the {\em formalism} he was exploring in his early work~\cite{db52} was central to quantum mechanics itself.  We have been able to confirm this in this paper where we have shown that the notions he was using  actually come directly from the energy-momentum tensor for the Schr\"{o}dinger field and had little to do with classical physics even though he had noticed that classical physics emerged as a limiting case.  This also shows that there is no need to derive the expression $p=\nabla S$ from some alternate fundamental principles which will enable us to regard this equation as a special defining equation for some new theory, {\em in addition to} the Schr\"{o}dinger equation.  Such an assumption leads to a very different theory, one that Bohm himself had rejected. 

\subsection{The Role of the Quantum Potential Energy}
 
Now let us return to examine equation (\ref{eq:qhj}) more closely.  As we have already pointed out, in a closed system this equation is an expression for the {\em local} conservation of energy, provided we regard the quantum potential energy, $QPE=-\hbar^2\nabla^2R(x,t)/2mR(x,t)$, as a new quality of energy present only in the quantum domain.  Its appearance had also been noted in previous work.  For example, it appeared in the hydrodynamical model introduced by Madelung ~\cite{em27}.  In this approach, the QPE could be considered as arising from some internal pressures in a `quantum fluid', the nature of which was left open.  In a very different context, Feynman~\cite{rf65} noticed how the QPE arose in a model of superconductivity he was exploring.  He called it a `quantum mechanical energy', but argued that in the superconductor itself it would be negligible and therefore did not develop the idea further.  

Generally the QPE has been regarded with suspicion and often rejected.  For example even in D\"{u}rr, Goldstein and Zanghi~\cite{ddsg96}, where one would expect to find support, they claim that there is a ``serious flaw in the quantum potential approach".  However the actual flaw was never clearly never spelt out.  Indeed we will show that there is no flaw and the QPE is essential to obtain a complete description of a quantum system.

Earlier Heisenberg~\cite{wh58} had regarded the QPE as being ``introduced  {\em ad hoc}" by Bohm, yet as  we have seen, equations (\ref{eq:Liouv}) and (\ref{eq:qhj}) are exactly equivalent to the Schr\"{o}dinger equation and  it is in equation (\ref{eq:qhj}) that the QPE appears.  It is not added, it arises as a consequence of the Lagrangian that Heisenberg himself had introduced.

Apart from being essential to ensure the conservation of energy, we will show how the QPE provides us with a way of describing quantum effects in terms of an actual evolving process.  Furthermore equation (\ref{eq:qhj}) helps us to understand how quantum mechanics reduces to classical mechanics in a suitable limit when the QPE becomes negligible and t here is a continuous transition from the quantum to the classical~\cite{bham95}.   It has been claimed that  the relation $\bm p=\nabla S$ is all that is necessary to obtain a complete description of quantum processes~\cite{ddsg96} but again this cannot be correct because, as we have seen, it involves only three of the components of the 16-component energy-momentum tensor and therefore cannot completely describe the unfolding process.  Below we will give some examples that illustrate why it is essential to include the QPE to obtain a complete description of a quantum system.

So far we have discussed the quantum behaviour of a single particle, and the QHJ equation (\ref{eq:qhj}) suggests that we can describe quantum processes using local fields and therefore every thing must be local.  However this is not correct even in the case of the single particle. But before developing this point, let us first look at a two-particle system where non-locality is clearly needed.  In fact it was the nature of the QPE that prompted Bell to propose his inequalities~\cite{jb71}. 

If two particles are described by an entangled wave function $\psi(x_1,x_2,t)$ then the QHJ becomes  
\begin{eqnarray}
\partial_t S+\frac{(\nabla_1S)^2}{2m_1}+\frac{(\nabla_{2} S)^2}{2m_2}-\frac{\hbar^2}{2m_1}\frac{\nabla_1^2R}{R}-\frac{\hbar^2}{2m_2}\frac{\nabla_2^2R}{R}+V=0.	\label{eq:2body}
\end{eqnarray}
Here $R=R(x_1,x_2,t)$ showing that this field depends on the position of both particles at the {\em same time} $t$, implying some form of non-locality.  Thus although the kinetic energy of each particle is `local', the QPE is non-local, appearing to `lock' the individual momenta of the two-particles together, not in a rigid way, but in a way that ensures the conservation of the total  energy of the pair.  They are locked into a common timeframe so that although the sub-systems retain their individuality, they are constrained by what Bohr calls the {\em individuality of the pair}.  This behaviour is an example of what has been called elsewhere `a long range order in momentum space', a feature found in liquid helium~\cite{fl54}.  Notice that if the wave function is a simple product, $\psi_1(x_1)\psi_2(x_2)$, then the equation splits into a pair of {\em local} equations, losing their ability to function as a pair.  

Thus here we see the difference between our approach and QFT itself.  In our approach, while appearing to keep everything local seems to succeed for the single particle, entangled multi-particle systems have a non-local component.  It is as if some of the energy becomes {\em delocalised}, leaving the remaining energy localised on each particle.  It is the delocalised energy that ensures that the pair are locked together so that they behave as a coherent pair.   Standard QFT avoids facing this question because they consider the global properties of the system, which actually integrates non-locality into the whole approach.

To see that this is so, suppose we consider the $n$-particle Schr\"{o}dinger equation
\begin{eqnarray*}
H|\Psi(t)\rangle=i\hbar\partial_t|\Psi(t)\rangle.
\end{eqnarray*}
Here $|\Psi\rangle$ is an eigenfunction of the number operator.  The state vector describing this situation can be written in the form 
\begin{eqnarray*}
|\Psi(t)\rangle=\frac{1}{\sqrt{ n!}}\int dx_1\dots\int dx_n\Psi^{(n)}(x_1\dots x_n;t)\psi^*(x_n)\dots\psi^*(x_1)|0\rangle.
\end{eqnarray*}
Here $\psi^*(x)=\sum_{i=1}^\infty\langle x|\lambda_i\rangle a^\dag_i$, $a^\dag_i$ being the creation operator and $|\lambda_i\rangle$ is a complete orthonormal set of one-particle states.  This formalism  has the great advantage in that it allows us to deal with an ensemble of particles whose number is not fixed. 
If we consider the two-body case we find after a little manipulation,  \begin{eqnarray*}
 \Psi^{(2)}(x_1,x_2,t)=\frac{1}{\sqrt{ 2!}}\langle0|\psi(x_1)\psi(x_2)|\Psi(t)\rangle.
 \end{eqnarray*}
 (For details, see Schweber~\cite{ss64}.) 
Thus the entanglement contained in $\Psi^{(2)}(x_1,x_2,t)$ is taken care of in the general formalism without explicitly drawing attention to its non-local aspects by formally introducing local {\em local} field operators $\psi(x_i)$ and then integrating over all space.  The commutation relation between local field operators at different points of space ensures `no signalling'  energy is involved.  Our approach\footnote{We are not claiming that our approach is an alternative to quantum field theory.  We are merely pointing out that our approach allows us to explore a different aspect of quantum theory.} allows us to examine the properties of this no signalling non-locality which, nevertheless locks  together each particle  in the entangled system.  Equation (\ref{eq:2body}) then shows how this translates into the wave function expressed in terms of the real variables $R(x_1,x_2,t)$ and $S(x_1,x_2,t)$.
 
\subsection{What is a `Particle' in the Quantum Domain?}

This raises the question as to what actually is the nature of the `particle' in the quantum domain, a question that has been raised recently in the context of standard QFT by Colosi and Rovelli~\cite{dccr09}.  If we insist that the `particle' is a classical object, some small `rock-like' object, then the QPE is totally inexplicable and looks like something artificial, even though it emerges from the same mathematics as the Schr\"{o}dinger equation.  We therefore   cannot ignore its appearance and we must surely explore its consequences.  

In actual fact the notion of a potential, classical or quantum, is a tricky one.  Its presence is only detected by the behaviour it induces in a particle.  Leech~\cite{jl65} in his classic book on classical mechanics,  regards a classical potential energy, $V$, as ``a fictitious quantity so defined that any change in its value compensates changes in kinetic energy".  He goes on to claim that its use is ``permissible if it helps describing the motion of the system".  

I would not call a potential `fictitious', but rather regard it as a potentiality which is only realised at the actual position of the particle. Often one can trace the source of a potential to, for example, a point source, like a charge.  But that is not always the case.  Coriolis and centrifugal forces do not arise in this way.  They appear as a result of observing phenomena from a rotating frame of reference.  Nevertheless the forces produce real effects.  In general relativity, the gravitational potential arises from the curvature of space and thus the geometry.  Again the notion of gravitational force appears because we map the movement of the object into the wrong space, in this case, into a Euclidean space.

Can we regard the force produced by the QPE as arising from geometric or topological feature of some deeper unfolding process?
Such a process may have a centre of energy but the explanation of its time evolution involves some deeper features that underlie the whole process. In this sense the `particle'  would appear more like a relativistic liquid drop than a rigid extended object~\cite{dbjv58}, but in saying that  we are not calling for a sub-quantum fluid of the type envisaged by Madelung~\cite{em27}.  Perhaps the QPE should be regarded as  a novel form of `internal' energy\footnote{ It is interesting to note that Souriau~\cite{js83} states that if we regard the Galilean group as a sub-group of the symplectic group $Sp(6)$, we can decompose the energy into two conserved quantities: the kinetic energy (of the centre of mass) and an internal energy.}.  This becomes more apparent if we write
\begin{eqnarray*}
\frac{1}{2m}(\nabla S)^2\leftrightarrow \frac{1}{2m}\left[(\nabla S)^2-\hbar^2\frac{\nabla^2R}{R}\right]
\end{eqnarray*}
implying that as the quantum property of the system emerges, there is a flow of energy between the kinetic and quantum potential energies, the actual flow being conditioned by the experimental environment.  In this way we have the possibility of a kind of wholeness that was advocated by Bohr~\cite{nb61}.

In classical physics, kinetic energy  has a well defined meaning.  It is the way a particle carries energy from one point to another.  It seems to correspond to something `real' and meaningful, so the appearance of the QPE in the energy flow seems strange, but suppose there exists an interchange between the KE and the QPE.  This suggestion is not as outrageous as it may seem at first sight.  Recall that in special relativity, the kinetic energy only appears as an approximation, emerging from the dynamical mass.  This 
shows that kinetic energy is not as `real' as it appears to be in the classical world.   An extended discussion of this point can be found in Holland~\cite{ph95}.

If there is an exchange of energy between the KE and the QPE, then clearly if we require the word `particle' to encompass this total process, so that the particle cannot be a classical `rock-like' entity as we have already remarked.  In the two-body problem described by equation (\ref{eq:2body}), the QPE carries a non-local feature, and can be regarded as an internal energy that binds the two centres of energy together.  Thus in this case, the internal energy involves {\em both} particles.  But remember Bell's inequality shows that we cannot think of the QPE as giving rise to some classical force acting between two `particles'.  Thus there must be a radically new process in which  some energy becomes `delocalised', yet continuing to form an irreducible whole.   
Something more radical is involved.  

One suggestion is that we should regard a `particle' as an invariant feature of some deeper underlying process that cannot be localised at a point in space-time yet retains a coherence that can be parameterised by some form of mean values such as, say, its centre of energy.  It is this parameter that traces out one of the solutions of the QHJ equation (\ref{eq:qhj}).  There is some evidence for a notion like this arising in the Pauli equation, but in the relativistic Dirac equation, two centres emerge as reported in Hiley and Callaghan ~\cite{bhbc2}.  These results, interesting a they are, will not be discussed further in this paper.  We will simply conclude by suggesting that our task is to find the precise nature of the energy transfer process.

 \section{Enter Weak Values}
 
 In order to move this discussion onto new ground, it would be good to have some experimental input.   Clearly the components of the energy-momentum  tensor involve energy and therefore should have experimental consequences.  Is it possible to measure the components of the energy-momentum tensor?   Obviously they cannot be measured directly as they are not eigenvalues of any operator and therefore cannot be found using a von Neumann-type measurement\footnote{However for a different view see Horton, Dewdney and Nesteruk~\cite{ghcdan00}.}.  Fortunately a new type of measurement has been introduced by Aharonov and Vaidman~\cite{yalv90} which depends on the notion of a `weak value' and it turns out that the components of the energy-momentum tensor are directly related to the weak values of the momentum operator.
 

To bring out this relationship, we start from the general definition of a weak value of an operator $\hat A$, namely,
\begin{eqnarray*}
\langle A\rangle_W=\frac{\langle\phi|\hat A|\psi\rangle}{\langle\phi|\psi\rangle}.
\end{eqnarray*}
Firstly we note that this is a complex number, therefore each weak value contains two real numbers.  
Secondly, the appearance of the denominator of $\langle\phi|\psi\rangle$ looks like a disaster-in-waiting because if the two states $|\phi\rangle$ and $|\psi\rangle$ are orthogonal, the weak value becomes infinite, but fortunately it is not actually a disaster.  

To see why this is so, let us  examine the general relation between the mean value of an operator $\hat A$ and the real part of its weak value $\langle A(\phi)\rangle_W$.  
In~\cite{ahys10, yasd05, bh12} it was shown that
\begin{eqnarray}
\langle\psi|\hat A|\psi\rangle=\sum_j\rho_\psi(\phi_j)Re\left[\frac{\langle\phi_j|\hat A|\psi\rangle}{\langle\phi_j|\psi\rangle}\right]=\sum_j\rho_\psi(\phi_j)Re\left[\langle A(\phi_j)\rangle_W\right].  	\label{eq:mvw}
\end{eqnarray}
Then as $Re[\langle A(\phi_j\rangle_W]\rightarrow \infty$, the probability density, $\rho_\psi(\phi_j)\rightarrow 0$, cancelling out the  contribution that this weak value makes to the overall expectation value.  However it should be noted that the
denominator strongly influences the magnitudes of the values, so there is the potentiality to amplify some processes, offering a new way to explore very weak effects.  For a detailed discussion of this feature in general see Flack and Hiley~\cite{rfbh13}.

Since, in this paper we are only interested in energy and momentum, we will confine our attention to weak values of the momentum operator at the point $x$.  In this case we put $\langle\phi|=\langle x|$, so that the denominator is simply $\psi(x)$.  No problem arises from the zeros in the wave function because they are physically uninteresting as the probability of finding a particle at such points is zero.  
Then a straight forward calculation shows that
\begin{eqnarray}
\langle P^j(x)\rangle_W=\frac{\langle x|\hat P^j|\psi\rangle}{\langle x|\psi\rangle}=\partial^j S(x,t)-i\hbar\partial^j\rho(x,t)/2\rho(x,t).	\label{eq:WVP}
\end{eqnarray}
We immediately see that the first term on the RHS is the real part of the weak value and, as we have already seen, it is related to the components of the momentum $T^{0j}$ through the relation $Re[T^{0j}(x,t)]=\rho(x,t) \partial^jS(x,t)$ with $\rho(x,t)=|\psi(x,t)|^2$.  
The second term, the imaginary part,  is known as the osmotic momentum, a term used by Bohm and Hiley~\cite{dbbh89} in their investigation of the relation of a stochastic version of a diffusion model for quantum mechanics that was extensively discussed by Nelson~\cite{en66}.  We will discuss the appearance and significance of the osmotic velocity later in section 5.

We can actually go further and evaluate the real part  of the weak value of $\hat P^2$.  This gives
\begin{eqnarray*}
\langle P^2\rangle_W=(\nabla S(x,t))^2-\hbar^2\nabla^2R(x,t))/R(x,t)
\end{eqnarray*}
showing that the real part of the weak value of the kinetic energy is simply the Bohm kinetic energy plus the QPE.  This equation immediately opens up the possibility of experimental access to the QPE provided we can find a way to make weak measurements. If this is possible then this opens up an experimental way of exploring the meaning of the QPE.

Before taking this particular investigation further here, let us look into the relation of some of the above results with standard quantum theory.  To this end let us pick up the discussion that is already present in Mackey~\cite{gm63}.  Let us form $\langle\psi|\hat P|\psi\rangle$ with $\langle x|\psi\rangle=Re^{iS/\hbar}$, then
\begin{eqnarray*}
\langle\psi|\hat P|\psi\rangle=\frac{\hbar}{i}\int\nabla(Re^{iS/\hbar})(Re^{iS/\hbar})d^3x\\=\int \rho\left[\frac{\hbar}{2i}\frac{\nabla\rho}{\rho}+\nabla S\right]d^3x.\hspace{0.6cm}	
\end{eqnarray*}
Note the appearance of the osmotic momentum $(\nabla \rho)/\rho$ in the intergrand.  However the contribution from this vanishes on integration because
 $\int\nabla \rho d^3x=0$
since $\rho\rightarrow0$ as $x\rightarrow\pm\infty$.  Finally we have
\begin{eqnarray}
\langle\psi|\hat P|\psi\rangle=\int\rho P_B d^3x=\int T^{0\mu} d^3x,	\label{eq:expPB}
\end{eqnarray}
which is identical to the momentum used in field theory to describe the moment of the particle as shown in equation (\ref{eq:momentum}).  Thus QFT does not pick up the osmotic part of the local momentum. Note that equation (\ref{eq:expPB}) can be written in the form
\begin{eqnarray*}
\langle\psi|\hat P|\psi\rangle=\int \rho(x)\Re[\langle P(x)\rangle_W]d^3x=\int \rho(x)P_B(x) d^3x
\end{eqnarray*}
showing how the Bohm momentum contributes to the mean value of the momentum revealed in a strong measurement.  Another example that we have used above is $\langle\psi|\hat P^2|\psi\rangle$.  This gives
\begin{eqnarray*}
\langle\psi|\hat P^2|\psi\rangle=\int\rho[(\nabla S)^2-\nabla^2R/R] d^3x=\int \rho(x)\langle P^2(x)\rangle_Wd^3x.
\end{eqnarray*}
Notice the QPE appearing in this equation, showing that it contributes to the mean of the operator $\hat P^2$.  Thus we see there is a very close relationship between conventional quantum field theory and the Bohm approach, much closer than one would first expect.

Interesting as these comparisons are, there is an even more important aspect, namely, the new technique of weak measurements now enables us to experimentally explore these relations in more detail~\cite{rfbh13}.  Under  a  suitable approximation, the weak measurement induces a phase change in the centre of mass wave function, which if coupled to the spin, changes the spin orientation.  Measuring the change in spin orientation enables the phase change to be measured indirectly thus giving a value of the Bohm momentum.  This technique has already been used to measure the transverse Poynting vector (i.e. the transverse Bohm momentum) of single photons in the interference region of a two-slit-type experiment.  Using these techniques Kocsis {\em et al} constructed energy flow lines that they interpreted as `photon trajectories' ~\cite{kas11}.

\section{Energy Eigenstates}

\subsection{Some Simple Examples of Energy Eigenstates}

In order to obtain some further insights as to how the QHJ equation (\ref{eq:qhj}) works in general, let us consider some simple examples of its use in stationary state situations.  Let us begin with a very simple quantum mechanical problem, namely the particle in a one-dimensional box with impenetrable walls.  As is well known, the $n$th energy eigenstate is given by 
\begin{eqnarray*}
\psi_n(x,t)=\sqrt{\frac{2}{a}}\sin\left(\frac{n\pi x}{a}\right)e^{\frac{-iE_nt}{\hbar}}.
\end{eqnarray*}
Here $a$ is the length of the box.  If we compare this with the standard form $\psi_n(x,t)=R_n(x,t)\exp[iS_n(x,t)/\hbar]$ we find
\begin{eqnarray}
\partial_xS_n=0\quad\mbox{and}\quad\partial_tS_n=-E_n.  \label{eq:En}
\end{eqnarray}
Thus the momentum flux, $T^{01}(x,t) = 0$, implying that, in this approach, the `particle' is at rest in every energy eigenstate in the box, so where is the energy?  In this state, the QHJ equation (\ref{eq:qhj}) becomes
\begin{eqnarray*}
\frac{\partial S_n}{\partial t}-\frac{\hbar^2}{2m}\left(\frac{\nabla^2R_n}{R_n}\right)=0.
\end{eqnarray*}
Using the last term in (\ref{eq:En}) gives
\begin{eqnarray*}
E_n=-\frac{\hbar^2}{2m}\left(\frac{\nabla^2R_n}{R_n}\right),
\end{eqnarray*}
so that the energy of the ground state is all QPE.   In fact we can quickly show this since 
\begin{eqnarray*}
Q_n=-\frac{\hbar^2}{2m}\nabla^2R_n/R_n\quad\mbox{with}\quad R_n=\sqrt{\frac{2}{a}}\sin\left(\frac{n\pi x}{a}\right).
\end{eqnarray*}
We find
\begin{eqnarray*}
Q_n=\frac{n^2\hbar^2\pi^2}{2ma^2}
\end{eqnarray*}
which will be immediately recognised as the energy eigenvalues of a particle trapped in a box.
Notice further that the energy is delocalised, giving the same value at any point in the `box'.

 The fact that the momentum flux, $T^{01}$, is zero can be interpreted in two ways.  First notice that the wave function can be written as
\begin{eqnarray*}
\sqrt{\frac{2}{a}}\sin\left(\frac{n\pi x}{a}\right)e^{\frac{iE_nt}{\hbar}}
=\sqrt{\frac{2}{a}}\left(\frac{1}{2i}\right)\left[e^{i(n\pi x/a-E_nt/\hbar)}-e^{-i(n\pi x/a+E_nt/\hbar})\right].
\end{eqnarray*}
This is normally taken to suggest the two components correspond to the particle running to the left and, {\em simultaneously}, running to the right, giving a net zero velocity. Or you can take the suggestion made by Bohm and Hiley~\cite{dbbh85} that the particle is actually at rest and all the energy of the bound state is trapped as QPE.  Then when the side of the box is removed suddenly, this quantum potential energy is all converted into particle momentum\footnote{I like to think of this as a  quantum KERS, where the acronym stands for Kinetic Energy Retrieval System.}
 with
\begin{eqnarray*}
p_B=\partial_xS_n=\pm\frac{n\pi \hbar}{a}\quad\mbox{and}\quad E_B=-\partial_tS_n=E_n.
\end{eqnarray*}

\subsection{Weak Values Again}

Finally a comment on a conclusion reached by Aharonov and Rolich~\cite{yadr05}, namely that a weak value can produce a `negative kinetic energy', is appropriate here.  The example they use is a particle trapped in a $\delta$-function well, which is essentially another example of a particle in a `box'.  The solution of the Schr\"{o}dinger equation in this case gives just one energy eigenstate
\begin{eqnarray*}
\psi(x)=\sqrt{\alpha}e^{-\alpha|x|}=R(x)\quad\mbox{with eigenvalue}\quad E=-\hbar^2\alpha^2/2m.
\end{eqnarray*}
We see immediately that the wave function is real so, like the particle in a box, the real part of the weak value of the momentum, the momentum flux, $T^{01}$, is zero, so the particle is again at rest.   
If we evaluate the real part of the weak value of the kinetic energy, we find
\begin{eqnarray*}
\langle P^2\rangle_W/2m=-\frac{\hbar^2}{2m}\nabla^2_xR(x)/R(x)=-\alpha^2\hbar^2/2m.
\end{eqnarray*}
Thus we see once again that the energy of the bound state is all QPE, the negative sign appearing because the particle is in a bound state.  Thus once again we have complete consistency.  Our approach does not require the particle to have a `negative kinetic energy', whatever that may mean.

\subsection{Stationary States in Atoms}

So far we have discussed rather artificial problems merely to show how the approach through equation (\ref{eq:qhj}) works.  Let us look at the simplest atom, namely, the hydrogen atom.  The ground state solution is
\begin{eqnarray*}
\psi_{1s}(r,\theta, \phi)=\frac{1}{\sqrt{\pi}}e^{br}e^{-iEt/\hbar}=Re^{iS/\hbar}\quad\mbox{with}\quad b=\frac{\mu e^2}{\hbar^2}.
\end{eqnarray*}
Since the phase of the wave function is only a function of $t$, $T^{0j}=0$, so that once again the particle is at rest.  In fact this result is true for all atomic $s$-state solutions of the Schr\"{o}dinger equation\footnote{For the Dirac equation the particle is not stationary in the ground state, a fact that we will discuss later.}, but it is not true for states with angular momentum.  In the latter we find the electron is moving to give the appropriate value of the angular momentum. 

 Again the behaviour of the electron in the $s$-state  seems rather unexpectedly `strange' but, of course, it must have zero angular momentum.  It is surely equally strange to understand how an electron could circulate around the nucleus with zero angular momentum.  The only way it can have zero angular momentum is to remain stationary. 

Let us now examine the behaviour of this state further and evaluate the remaining term in the QHJ, namely the quantum potential energy We find
\begin{eqnarray*}
Q=-\frac{\hbar^2}{2\mu}\frac{\nabla^2R}{R}=\frac{e^2}{r}-\frac{\mu e^4}{2\hbar^2}
\end{eqnarray*}
so that equation (\ref{eq:qhj}) gives
\begin{eqnarray*}
E=Q +V=Q-\frac{e^2}{r}=-\frac{\mu e^4}{2\hbar^2}.
\end{eqnarray*}
Thus we see that the quantum potential energy minus the Coulomb energy gives us the energy of the ground state.  Notice once again the QPE is delocalised, being the same at every point for which $|\psi(x,t)|^2 \ne 0$. Notice also that if we form
\begin{eqnarray*}
\nabla E=\nabla Q+\nabla V,
\end{eqnarray*}
we see that the force calculated from the quantum potential  balances exactly the Coulomb force, so that dynamically everything seems to fit with the concept of a point particle even though it does not fit well with our concepts of a {\em classical} particle.

Apart from the `unreasonableness' of the stationary particle, it is well known that if in the hydrogen atom, the electron is replaced by a muon, the half-life of the muon is dilated in the ground state. This fact surely indicates that the muon is in motion contrary to the results we have just found.  In this model replacing the electron with a muon merely changes the ground state energy, the phase of the wave function remains independent of $r$ so that $T^{0j}=0$ is again the result.  This suggests that something is seriously wrong\footnote{ As a historical point of interest, it was the stationary nature of the electron in these cases that drew the strongest criticism from both Einstein~\cite{ae53} and Heisenberg~\cite{wh58}.  The many developments, both theoretical and experimental, since then suggest their conclusions were based on metaphysical assumptions that are now untenable. }.

However before rejecting the whole model on this argument, one should realise that time dilation is a relativistic effect, whereas any approach using the non-relativistic Schr\"{o}dinger equation might not pick up something that is perhaps contained in the relativistic Dirac equation.  This, in fact, turns out to be the case.  If one uses the relativistic generalisation of the velocity discussed in Bohm and Hiley~\cite{dbbh93}
\begin{eqnarray}
v^j=\frac{\bar\psi\gamma^j\psi}{|\psi|^2}	\label{eq:vdirac}
\end{eqnarray}
and uses the ground state solution of the Dirac equation for the hydrogen atom, we find that the electron is in fact moving.  Furthermore if we examine the non-relativistic limit of the expression (\ref{eq:vdirac}) in the ground state, we find $v^j=0$, in agreement with the result obtained from the Schr\"{o}dinger equation. We will not discuss the details of this result here as they can be found in Hiley~\cite{bhvdirac}.

\section{Towards a Deeper Theory}

\subsection{The Need to go deeper}

As we have seen from the comments of Einstein~\cite{ae53} and Heisenberg~\cite{wh58}, these results do not encourage confidence and therefore have sometimes been referred to as `bizarre'.  However to label something `bizarre' does not necessarily imply it is wrong.  Quantum phenomena are `bizarre' since they are very different from classical phenomena.  Therefore one should expect any explanation to be unexpected.   On the other hand, the structure of the QHJ equation (\ref{eq:qhj}) is so temptingly close to classical physics, that one is tempted to try to interpret the results using classical intuition. Unfortunately this conclusion may well be wrong because we are dealing with something very different, but what?

The mathematical structure that we explored above is essentially based on the Schr\"{o}dinger formalism and the Lagrangian was chosen so that the Euler-Legrange equations would, in fact, give the Schr\"{o}dinger equation. But where does the Schr\"{o}dinger equation actually come from in the first place?  Examination of Schr\"{o}dinger's original papers shows it was arrived at using a heuristic argument and was not derived from any set of underlying principles.  Schr\"{o}dinger had noticed the close relationship between geometric (ray) optics and Hamilton's mechanics and was looking for what he called ``a Hamiltonian undulatory mechanics".

At the time Schr\"{o}dinger was searching for his ``undulatory mechanics", symplectic geometry, the geometry that underlies Hamiltonian mechanics, was not sufficiently well developed to enable him to exploit this rich geometric structure.  Since the `60s there has been a considerable volume of work done to understand symplectic geometry, which is very different from Euclidean or Minkowski geometry.  However one can find a good account of this rich structure accessible to physicists in the books of de Gosson~\cite{mdg01, mdg10}.  An understanding of these new insights has led de Gosson and Hiley~\cite{mdgbh11} to show how the Hamiltonian flow can be `lifted' onto the covering group of the symplectic group, producing a flow that is described by the Schr\"{o}dinger equation.  This confirms the results derived by Guillemin and Sternberg~\cite{vgss84} who show that the Schr\"{o}dinger equation resides in the covering space of a symplectic space.  This has opened up a new area of investigation that throws considerable light on this deeper structure as we will show below.

\subsection{Consequences of Non-commutativity}

To begin our discussion, let us return to the discussion of the weak value of the momentum operator.  As we have seen, the real part of the weak value of the momentum is the  momentum flow, $P_B$, while the imaginary part of this value suggests a new form of momentum which we have called the `osmotic momentum'.  Bohm and Hiley~\cite{dbbh89} came across such a momentum when exploring a stochastic approach to quantum mechanics, a notion that was discussed earlier by  Nelson~\cite{en66}. 

When exploring a stochastic approach to quantum mechanics, Nelson~\cite{en85} argued that there was a need to generalise the notion of a derivative because ``Nature operates on a different scheme in which the past and the future are not on an equal footing".  To this end he defined a `forward' derivative
\begin{eqnarray}
Dx(t)=\lim_{\Delta t\rightarrow0+}E_t\frac{x(t+\delta t)-x(t)}{\Delta t}, \label{eq:fd}
\end{eqnarray}
where $E_t$ denotes the conditional expectation given $x(t)$, and a 'backward' derivative
\begin{eqnarray}
D_*x(t)=\lim_{\Delta t\rightarrow0+}E_t\frac{x(t)-x(t-\delta t)}{\Delta t}.  \label{eq:bd}
\end{eqnarray}
If $x(t)$ is differentiable, then $Dx(t)=D_*x(t)=dx/dt$, but if not, we must distinguish these two derivatives. 

Nelson goes on to use these derivatives to construct what he called the mean {\em forward} velocity, $\bm b(\bm x(t),t)$,
\begin{eqnarray}
D\bm x(t)=\bm b(\bm x(t),t)	\label{eq:fv}
\end{eqnarray}
and mean {\em backward} velocities, $\bm b^*(\bm x(t),t)$,
\begin{eqnarray}
D_*\bm x(t)=\bm b_*(\bm x(t),t).		\label{eq:bv}
\end{eqnarray}
By comparing these variables with Einstein's theory of Brownian motion, we find $\bm v=(\bm b+\bm b_*)/2$ is the current velocity and $\bm v_0=(\bm b-\bm b_*)/2$ is the osmotic velocity.
Applying these ideas to a stochastic approach to quantum mechanics, Bohm and Hiley~\cite{dbbh89} obtained the following results:
\begin{eqnarray}
\bm v=(\bm b+\bm b_*)/2=\nabla S/m=\bm v_B		\label{eq:vb}
\end{eqnarray}
and 
\begin{eqnarray}
\bm v_0=(\bm b-\bm b_*)/2=\frac{\hbar}{2m}\frac{\nabla \rho}{\rho}=\frac{\hbar}{m}\frac{\nabla R}{R}.		\label{eq:vo}
\end{eqnarray}
These two factors appear in the weak value of the momentum shown in equation (\ref{eq:WVP}).

If one returns to the earlier work of Madelung~\cite{em27}, the appearance of this additional momentum offered support  to the suggestion that the quantum particle is in fact subjected to fluctuations in some novel form of `sub-quantum medium'~\cite{lb64, en66}.  However this approach suffers from the same criticism used against the earlier notion of a {\em luminiferous aether}. Einstein's way out of this difficulty was to propose we take the Maxwell field itself as a basic entity and not look for any {\em mechanical} explanation in terms of some underlying medium.  Let us follow Einstein's advice in this case, formally adopting these two velocities in their own right and simply drop an explanation in terms of some mechanical sub-quantum medium.
  
  It should be noted that these generalised derivatives also appear in the treatment of the Pauli and Dirac particles~\cite{bhbc12}.  This means that these results, together with similar relationships between the energy-momentum tensor and the Bohm momentum and Bohm energy, are quite general and not confined to the particular form of the  Schr\"{o}dinger equation.  
  
The appearance of backward and forward derivatives means that we should distinguish between operators operating from the left and those operating from the right.  In other words, in quantum mechanics, we must distinguish between left and right translations.  To explore the consequences of this distinction,
let us take the Schr\"{o}dinger equation and its dual given in equation (\ref{eq:ss}) and write them in a more transparent form
\begin{eqnarray}
	i\hbar{\overrightarrow\partial_t }|\psi(t)\rangle={\overrightarrow H}|\psi(t)\rangle\quad\mbox{and}\quad -i\hbar\langle\phi(t)|{\overleftarrow\partial_t}=\langle\phi(t)|{\overleftarrow H}.		\label{eq:leftright}
\end{eqnarray}
 
 If one is only interested in calculations, then it looks as if we are introducing unnecessary complications.  However we are interested in the complete mathematical structure, including its subtleties, as we believe they provide clues to some new physical insights.  To this end we argue that to capture the full implications of quantum processes, we need a more general mathematical structure, a bimodule structure~\cite{ia72}.  We will show that the usual configuration space is but a projection of the full structure which allows us to treat the formalism as a one-side left module or, in more familiar terms, a left vector space, which is the traditional approach that most physicists are comfortable with.

The distinction between left and right translations has been hinted at in the physics literature already, but its relevance does not seem to have been recognised in general.   Feynman~\cite{rf48} is an exception. In setting up the sum-over-paths approach, he noted that $|\psi(t)\rangle$ contains information coming from the past $t_i<t$, and does not depend in anyway on what will happen in the future. 
On the other hand $\langle\phi(t)|$ ``characterizes the experiences, or experiments to which the system {\em is to be subjected} "--if you like, it is a symbol signifying an ``anticipation of the future".  Feynman then takes $\langle\phi(t)|\psi(t)\rangle$ as the probability amplitude for a system in the state $|\psi\rangle$  ending up in a future state $|\phi\rangle$.  

However one can also write
$|\psi(t)\rangle\langle\phi(t)|$ which can be used to represent a transition operator. In this way Feynman has given a physical reason for us to distinguish between the left and right translations, but develops it only as  a calculation tool.     Sometimes $\langle\phi(t)|$ has been interpreted as ``information coming from the future" but perhaps this interpretation is a step too far.

Let us look at the bimodule structure motivated from a physical point of view.  To this end we refer to Dirac~\cite{pd30}, who in a very early paper, had already drawn attention to the difference between the left and right translations appearing in equation (\ref{eq:leftright}).  He pointed out that when we deal with a single pure state, we describe it by $|\psi(t)\rangle$ and use the Schr\"{o}dinger equation to determine its time evolution.  On the other hand when we are dealing with an ensemble of states, we use the density matrix $\rho$ satisfying the equation of motion
\begin{eqnarray}
i\hbar\dot\rho=H\rho-\rho H.		\label{eq:dmtime}
\end{eqnarray}
However there may be intermediate situations where it is convenient to combine the advantages of both equations.  That is, if we write $\rho(t)=|\psi(t)\rangle\langle\phi(t)|$ then
\begin{eqnarray}
i\hbar\frac{\partial\rho(t)}{\partial t}=i\hbar\frac{\partial}{\partial t}(|\psi(t)\rangle\langle \phi(t)|)=i\hbar\left[\left(\frac{\partial|\psi(t)\rangle}{\partial t}\right)\langle \phi(t)|+|\psi(t)\rangle\left(\frac{\partial\langle\phi(t)|}{\partial t}\right)\right]\nonumber\\
=\left(\overrightarrow H|\psi(t)\rangle\right)\langle\phi(t)|-|\psi(t)\rangle\left(\langle\phi(t)|\overleftarrow H\right)=H\rho(t)-\rho(t) H.\hspace{1.35cm}\label{eq:le1}
\end{eqnarray}
Dirac has in mind a situation where $\psi$ is represented by a rectangular $n\times m$ matrix, while $\phi$ is represented by an $m\times n$ matrix. If $m >>n$, then it is easier to solve equation (\ref{eq:le1}) than the full Schr\"{o}dinger equation involving an $n\times n$ matrix.

Our concern is not whether one approach is simpler to apply than another, important though that may be.  We want to understand the overall mathematical structure that distinguishes left translation from right translations and what it implies for the physics so let us look more deeply into equation (\ref{eq:le1}).  

This equation  can be constructed more simply from the first equation in (\ref{eq:leftright}) by writing
\begin{eqnarray}
i\hbar\left({\overrightarrow\partial_t}|\psi\rangle\right)\langle\phi|=\left({\overrightarrow H}|\psi\rangle\right)\langle\phi| \label{eq:sden}
\end{eqnarray}
and then writing the second equation in (\ref{eq:leftright}) in the form
\begin{eqnarray}
 -i\hbar|\psi\rangle\left(\langle\phi|{\overleftarrow\partial_t}\right)=|\psi\rangle\left(\langle\phi|{\overleftarrow H}\right).	\label{eq:csden}
\end{eqnarray}
Subtracting  equation (\ref{eq:sden}) from  (\ref{eq:csden}) gives equation (\ref{eq:le1}) so that we are back to Dirac's equation.  However we can do something different by adding equation (\ref{eq:sden}) to (\ref{eq:csden}) and obtain the equation
\begin{eqnarray}
i\hbar\left[\left({\overrightarrow\partial_t}|\psi\rangle\right)\langle \phi|-|\psi\rangle\left(\langle\phi|{\overleftarrow\partial_t}\right)\right]\hspace{5cm}\nonumber\\
=\left(\overrightarrow H|\psi\rangle\right)\langle\phi|+|\psi\rangle\left(\langle\phi|\overleftarrow H\right)=H\rho+\rho H	\label{eq:plus}
\end{eqnarray}
where we have written  $\rho=|\psi\rangle\langle\phi|$. If we follow the usual convention and write
\begin{eqnarray*}
\left({\overrightarrow \partial_t}|\psi\rangle\right)\langle \phi|-|\psi\rangle\left(\langle\phi|{\overleftarrow\partial_t}\right):= |\psi\rangle{\overleftrightarrow\partial_t}\langle\phi|,
\end{eqnarray*}
we can write equation (\ref{eq:plus}) in a simpler form 
\begin{eqnarray}
i\hbar |\psi\rangle{\overleftrightarrow\partial_t}\langle\phi|=H\rho+\rho H.	\label{eq:acom}
\end{eqnarray}
If we take this together with
\begin{eqnarray}
i\hbar\partial_t\rho=H\rho-\rho H.		\label{eq:com}
\end{eqnarray}
 we have replaced the Schr\"{o}dinger equation and its dual  by two different equations (\ref{eq:acom})  and (\ref{eq:com}).
These two equations were first obtained  by Brown and Hiley~\cite{mbbh00} using a different argument.  

\subsection{Meaning of the Two Equations}

The meaning of equation (\ref{eq:com}) can be brought out by first writing $\rho=|\psi(t)\rangle\langle\psi(t)|$ and then using the projection operator $\Pi_x=|x\rangle\langle x|$ and the Hamiltonian $H=-\frac{\hbar^2}{2m}\nabla^2 +V$, we find
\begin{eqnarray*}
\frac{\partial\rho}{\partial t}+\nabla.\left(\rho\frac{\nabla S}{m}\right)=0
\end{eqnarray*}
where we have written $\psi=Re^{iS/\hbar}$ with $R^2=\rho$. We immediately recognise this as being the quantum Liouville equation given in equation (\ref{eq:Liouv}).  Clearly in this context the equation guarantees the conservation of probability.  

The meaning of equation (\ref{eq:acom}) is far from clear as it stands, but once again if we project it into the $x$-representation we find it reads
\begin{eqnarray*}
\frac{\partial S}{\partial t}+\frac{(\nabla S)^2}{2m}-\frac{\hbar^2}{2m}\left(\frac{\nabla^2R}{R}\right)+V=0
\end{eqnarray*}
which is, of course, identical to the quantum Hamilton-Jacobi equation (\ref{eq:qhj}) showing that the QPE only appears as a result of using the projection operator $\Pi_x$.  

However, as shown by Brown and Hiley~\cite{mbbh00}, we need not restrict ourselves to one specific representation; we can choose the projection operator $\Pi_p=|p\rangle\langle p|$ to find a pair of equations for the $p$-representation.  The first will correspond to the conservation of probability in momentum space, while the  second is a quantum Hamilton-Jacobi-type equation in momentum space.  This will contain a term for the QPE, but now in momentum space so that we could find a set of `trajectories'  but this time in the momentum space. This restores the $x-p$ symmetry that Heisenberg~\cite{wh58} claimed was missing from the original Bohm approach.  The Bohmian mechanics of D\"{u}rr {\em et al} chooses the position representation, claiming the momentum representation has no meaning.

For further discussion of the momentum representation and its consequences see Brown and Hiley~\cite{mbbh00} and Hiley~\cite{bh13}.   It is not necessary to restrict the projections to just these two representations.  A continuum of intermediate representations are possible that correspond to fractional Fourier transformations. A detailed discussion of this approach will be found in Brown~\cite{mb04}. 
 
\subsection{The von Neumann Algebra}

The way we have introduced the two defining non-commuting equations  (\ref{eq:com}) and (\ref{eq:acom}) can, at best, be regarded as heuristic.  I want now to show how a similar pair of equations arise from a non-commuting algebra\footnote{Here we are focussing our attention only on type-I von Neumann algebras.} originally introduced by von Neumann~\cite{vn31}.  In this approach what is normally called the density operator, $\rho$, rather than the wave function plays a fundamental role in defining every physical process in the quantum domain.  We will give a new meaning to this operator because we are applying it to a single particle.  It will show that the evolution of this `particle' requires a non-local description and that this non-locality is reflected in the symplectic structure that physicists refer to as `phase space'.

  von Neumann, following Weyl~\cite{hw28}, starts with a pair of bounded operators, $U(\alpha)=e^{i\alpha {\widehat P}}$ and $V(\beta)=e^{i\beta{\widehat X}}$.  These operators can be thought of as translations in position space and momentum space respectively.  Since the operators $\hat P$ and $\hat X$ no longer commute, the product $U(\alpha)V(\beta)$ is non-commutative and defined by
\begin{eqnarray}
U(\alpha)V(\beta) = e^{i\alpha\beta}V(\beta)U(\alpha),	\label{eq:Weyl}
\end{eqnarray}
together with
\begin{eqnarray}
U(\alpha)U(\beta)=U(\alpha + \beta);  \hspace{0.5cm}  V(\alpha)V(\beta)=V(\alpha + \beta). \nonumber
\end{eqnarray}	
Equation (\ref{eq:Weyl}) implicitly contains  the well known relation 
\begin{eqnarray*}
[\hat x, \hat p]=i\hbar.
\end{eqnarray*}							
von Neumann then defines an operator
\begin{eqnarray}
\widehat S(\alpha,\beta)=e^{i(\alpha\widehat P+\beta \widehat X)}=e^{-i\alpha\beta/2}U(\alpha)V(\beta)=e^{i\alpha\beta/2}V(\beta)U(\alpha)  \nonumber
\end{eqnarray}
and proves that the operator $\widehat S(\alpha, \beta)$ can  be used to define a bounded operator $\hat A$ on a Hilbert space  through the relation
\begin{eqnarray}
\hat A=\int\int a(\alpha, \beta)\widehat S(\alpha,\beta)d\alpha d\beta.
\label{eq:symA}				
\end{eqnarray}
Here $a(\alpha,\beta)$ is a function on a Schwartz space spanned by two variables $\alpha$ and $\beta$ in $\mathbb R^{2N}$.  In this space von Neumann introduced a non-commutative multiplication defined by
 \begin{eqnarray}
 a(p,x)\star b(p,x)=\int\int e^{2i(p\eta-x\xi)}a(p-\xi,x-\eta)\bar b(\xi,\eta)d\xi d\eta,			\label{eq:starvn}		
 \end{eqnarray}
 where
 \begin{eqnarray*}
 b(p,x)=\int\int e^{2i(p\eta-x\xi)}\bar b(\xi,\eta)d\xi d\eta.
 \end{eqnarray*}
What von Neumann has constructed here is a non-commutative symplectic space over $\mathbb R^{2N}$, which the physicist will recognise as some form of non-commutative phase space once we follow Moyal~\cite{jm49} and make the identification $\alpha=p$ and $\beta=x$.  Then it is possible to show that equation (\ref{eq:starvn}) for $(p,x)$ becomes
\begin{eqnarray*}
x\star p -p\star x=i\hbar.
\end{eqnarray*}
This shows how the Heisenberg defining relation appears in this symplectic phase space that has been constructed. For the mathematical details of symplectic geometry see de Gosson \cite{mdg12}, \cite{mdg01}, while a rigorous discussion of the Moyal algebra will be found in Gracia-Bondia and V\'{a}rilly~\cite{gbjv88}.  Some excellent work examining this algebra that is more ameniable to the physicist can be found in Curtright, Fairlie and Zacos~\cite{tcdf98} and Zacos~\cite{cz02}.

The star-product is called the Moyal star product  since it was Moyal who showed it could be written as a formal series
\begin{eqnarray}
a(p,x)\star b(p,x)= a(p,x) \exp\left[\frac{i\hbar}{2}\left(\overleftarrow{\frac{\partial}{\partial x}}\overrightarrow{\frac{\partial}{\partial p}}-\overrightarrow{\frac{\partial}{\partial x}}\overleftarrow{\frac{\partial}{\partial p}}\right)\right]b(p,x).  \label{eq:star}		
\end{eqnarray}
The star product enables us to define two types of bracket; the Moyal bracket defined by
\begin{eqnarray*}
\{a,b\}_{MB}=\frac{a\star b-b\star a}{i\hbar}=2a(p,x)\sin\left[\frac{\hbar}{2}\left(\overleftarrow{\frac{\partial}{\partial x}}\overrightarrow{\frac{\partial}{\partial p}}-\overrightarrow{\frac{\partial}{\partial x}}\overleftarrow{\frac{\partial}{\partial p}}\right)\right]b(p,x),
\end{eqnarray*}
and the Baker bracket \cite{bak}  (or Jordan product) defined by
\begin{eqnarray*}
\{a,b\}_{BB}=\frac{a\star b+b\star a}{2}=a(p,x)\cos\left[\frac{\hbar}{2}\left(\overleftarrow{\frac{\partial}{\partial x}}\overrightarrow{\frac{\partial}{\partial p}}-\overrightarrow{\frac{\partial}{\partial x}}\overleftarrow{\frac{\partial}{\partial p}}\right)\right]b(p,x).
\end{eqnarray*}
Clearly the Moyal bracket replaces the quantum operator commutation relations $[\hat A, \hat B]$.  

The nature of the $\star$-product means that the Moyal bracket will, in general, be a power series in $\hbar$.  If we  retain only the terms to $O(\hbar)$, we find 
\begin{eqnarray*}
\mbox{Moyal bracket}\rightarrow \mbox{Possion bracket}.
\end{eqnarray*}
 While in the case of the Baker bracket we find
\begin{eqnarray*}
\mbox{Baker bracket}\rightarrow \mbox{simple commutative product}.
\end{eqnarray*}
Thus we find that the Moyal approach contains classical physics as a limiting case.  Notice also that so far the wave function has not made an appearance, so let us now turn to see how von Neumann and Moyal introduce the wave function.

\section{A Statistical Phase Space Theory}

Equation (\ref{eq:symA}) shows how any bounded operator on a Hilbert space can be replaced by the non-commutative algebra defined by the functions of the type $a(p,x)$, but as yet we have given no meaning to the operator $\widehat S(p,x)$.  To see what this symbol means let us form the expectation value	
\begin{eqnarray}
\langle \psi(t)|\widehat A|\psi(t)\rangle = \int\int a(p,x)\langle \psi(t)|\widehat S(p,x)|\psi(t)\rangle dp dx\nonumber
\\=\int\int a(p,x) F_\psi(p,x,t)dpdx. \hspace{1.3cm}      \label{eq:expA}		
\end{eqnarray}
Now let us compare this with the standard approach to the expectation value of an operator $\hat A$ which is written in the form 
\begin{eqnarray*}
\langle\psi(t)| \widehat A|\psi(t)\rangle &=&\int\int\langle \psi(t)|x'\rangle\langle x'|\hat A|x''\rangle\langle x''|\psi(t)\rangle dx'dx''\\
&=&\int\int A(x',x'')\rho_\psi(x',x'',t)dx'dx''
\end{eqnarray*}
where we have  written $\rho_\psi(x',x'',t)=\langle x''|\psi(t)\rangle\langle \psi(t)|x'\rangle$ in the form of a two-point density matrix.  
Let us now change to the coordinates 
\begin{eqnarray*}
x'=x-y/2;\quad\quad x''=x+y/2
\end{eqnarray*}
so that
\begin{eqnarray}
\langle \psi(t)|\widehat A|\psi(t)\rangle=\int\int \langle x-y/2|\hat A|x+y/2\rangle\rho_\psi(x-y/2,x+y/2,t)dxdy	.	\label{eq:xy}
\end{eqnarray}
Define $F_\psi(p,x,t)$ through the relation 
\begin{eqnarray*}
\rho_\psi(x-y/2,x+y/2,t)=\int F_\psi(p,x,t)e^{-iyp}dp.
\end{eqnarray*}
Substituting this expression into equation (\ref{eq:xy}) and rearranging, we find
\begin{eqnarray*}
\langle\psi(t)| \widehat A|\psi(t)\rangle=\int\int\left[\int\langle x-y/2|\hat A|x+y/2\rangle e^{-iyp}dy\right]F_\psi(p,x,t)dpdx.
\end{eqnarray*}
Defining the square bracket to be $a(p,x)$, we find
\begin{eqnarray}
\langle\psi(t)| \widehat A|\psi(t)\rangle=\int\int a(p,x)F_\psi(p,x,t)dpdx   \label{eq:Maverage}
\end{eqnarray}
which is identical to equation (\ref{eq:expA}).  Thus we have identified $F_\psi(p,x,t)$ as the Fourier transform of the two-point density matrix in configuration space.  Since $p$ and $x$ are the coordinates of a non-commutative symplectic phase space, this equation suggests that  $F_\psi(p,x,t)$  can be regarded as a density matrix in that space~\cite{dbbh81}. Thus we should regard equation (\ref{eq:Maverage}), not as an attempt to return to classical ideas, but simply as a different way of writing quantum expectation values in the same spirit that this value can be rewritten for weak values as demonstrated by equation (\ref{eq:mvw}).

 In this symplectic phase space, the labels $(p,x)$ turns out not to be values of the momentum and position of a particle, but can be shown to be the mean momentum and mean position of a cell in phase space.  In other words, the energy of a single particle is not localised at a point, but extends over an extended region in phase-space. In other words, in this theory non-commutativity builds in non-locality from the very beginning.  The finite region  is called the quantum blob of de Gosson~\cite{mdg12} where more details can be found.  Thus $F_\psi(p,x,t)$ describes the state of a {\em region} in phase space, not the state of a point particle.

\subsection{Time Development in Phase Space}

Let us now examine the time evolution of $F_\psi(p,x,t)$.  Since  our algebra is non-commutative, we again must distinguish between left and right translations, that is between $H(p,x)\star F_\psi(p,x,t)$ and $F_\psi(p,x,t)\star H(p,x)$, where $H(p,x)$ is the Hamiltonian.  This means that again we have two equations for the time development,
\begin{eqnarray}
H(x,p)\star F_\psi(x,p,t)=i(2\pi)^{-1}\int e^{-iyp}\psi^*(x-y/2,t)\overrightarrow\partial_t\psi(x+y/2,t)dy  \label{eq:LM}	
\end{eqnarray}
and
\begin{eqnarray}
F_\psi(x,p,t)\star H(x,p)=-i(2\pi)^{-1}\int e^{-iyp}\psi^*(x-y/2,t)\overleftarrow\partial_t\psi(x+y/2,t)dy.  \label{eq:RM}	
\end{eqnarray}
If we subtract these two equations, we immediately obtain 
\begin{eqnarray}
i\partial_tF_\psi(x,y,t)=\{H,F\}_{MB},		\label{eq:mbt}
\end{eqnarray}
which has a strong resemblance with equation (\ref{eq:com}).  
We can also add the two equations (\ref{eq:LM}) and (\ref{eq:RM}) to find
\begin{eqnarray}
\{H,F\}_{BB}=i(2\pi)^{-1}\int e^{-iy p}\left[\psi^*(x-y/2,t)\overleftrightarrow\partial_t\psi(x+y/2,t)\right]dy, 	\label{eq:bb}
\end{eqnarray}  
where we have introduced a condensed notation which is well known in quantum field theory\footnote{The distinction between left and right multiplication is necessary even in conventional quantum field theory when one deals with the Pauli and Dirac particles.  The double arrow symbol (\ref{eq:2way}) is used for the energy term, for example, in the Lagrangian for the Dirac field \cite{pr81}.  It is therefore not surprising that equation (\ref{eq:bb}) involves energy.}, viz,
\begin{eqnarray}
\psi^*\overleftrightarrow \partial_t \psi=\psi^*(\partial_t\psi)-(\partial_t\psi^*)\psi.		\label{eq:2way}	
\end{eqnarray}
We can quickly get an idea as to what the RHS of equation (\ref{eq:bb}) means by choosing an energy eigenstate, $\psi(x, y,t)=\phi(x, y)e^{iEt}$.  We  find
\begin{eqnarray*}
i(2\pi)^{-1}\int e^{-iyp}\left[\psi^*(x-y/2,t)\overleftrightarrow\partial_t\psi(x+y/2,t)\right]dy =-2EF_\psi(p,x,t).
\end{eqnarray*}
So clearly the Baker bracket has something to do with the energy of the system.  Therefore let us condense the notation by writing
\begin{eqnarray*}
i(2\pi)^{-1}\int e^{-iyp}\left[\psi^*(x-y/2,t)\overleftrightarrow\partial_t\psi(x+y/2,t)\right]dy=-{\cal E}(p,x,t),
\end{eqnarray*}
so that we can write equation (\ref{eq:bb}) in the form
\begin{eqnarray}
{\cal E}(p,x,t)+ \{H,F\}_{BB}=0.		\label{eq:bbt}
\end{eqnarray}
This equation has a striking resemblance to equation (\ref{eq:acom}).

Notice also that if we use  two different energy eigenfunctions, by replacing $\psi^*(x,y,t)$ by $\psi_1^*(x,y,t)$ and writing $\psi_1^*(x,y,t)=\phi_1^*(x,y)e^{-iE_1t}$ then the Baker bracket measures the mean energy of the two eigenstates.  Such an equation was first introduced by Dahl \cite{jpd83}  to supplement the Liouville equation in order to obtain a complete specification of the energy eigenstates  of molecules \cite{pcfz83}.

The full significance of equation (\ref{eq:bb}) is still not obvious from the form of the RH term, but a further insight can be found by going to the limit $O(\hbar^2)$.  After some work, including writing $\psi(x,t)=R(x,t)\exp[iS(x,t)/\hbar]$, we find
\begin{eqnarray*}
\{H,F\}_{BB}=-2(\partial_tS) F+O(\hbar^2).
\end{eqnarray*}
From the definition of the Baker bracket, we find that in this limit equation (\ref{eq:bb}) becomes
\begin{eqnarray*}
\frac{\partial S}{\partial t}+H=0.
\end{eqnarray*}
This will immediately be recognised as the classical Hamilton-Jacobi equation. Thus equation (\ref{eq:bb}) is the quantum generalisation of the classical Hamilton-Jacobi equation hence the reason for calling equation (\ref{eq:qhj}) the quantum Hamilton-Jacobl equation.

Now let us return to the operator formalism and the two equations (\ref{eq:com}) and (\ref{eq:acom}).  We immediately see a correspondence emerging between these equations and the operator equations (\ref{eq:mbt})  and (\ref{eq:bbt}), namely
\begin{eqnarray*}
\{H,F\}_{MB}\quad\leftrightarrow\quad [\hat H,\hat\rho]_-,
\end{eqnarray*}
while the Baker bracket is the Moyal algebraic equivalent of the anti-commutator 
\begin{eqnarray*}
\{H,F\}_{BB}\quad\leftrightarrow\quad [\hat H, \hat \rho]_+.
\end{eqnarray*}

The operator equation is the analogue of equation (\ref{eq:com}) which is the quantum Liouville equation
\begin{eqnarray*}
\frac{\partial \hat\rho_\psi(x,t)}{\partial t}+[\hat H(x), \hat\rho_\psi(x,t)]_-=0.
\end{eqnarray*}
While equation (\ref{eq:acom}) has the operator analogue equation
\begin{eqnarray}
2\partial_tS\hat\rho+[\hat \rho,\hat H]_+=0.		\label{eq:bb0}	
\end{eqnarray}

\subsection{From Phase Space to Configuration Space}

We have used the word `corresponds' in comparing the pair of equations (\ref{eq:com}) and (\ref{eq:acom}) and the pair of equations (\ref{eq:mbt}) and (\ref{eq:bbt}), but they are qualitatively different in the sense that the former are operator equations which can be represented in configuration space and the latter are equations in a non-commutative symplectic phase space.  Thus we must project equations (\ref{eq:mbt}) and (\ref{eq:bbt}) from the phase space to the configuration space which is a subspace to compare like with like.

The way we do this is by constructing conditional probabilities.  For example, Moyal~\cite{jm49} constructs a conditional expectation of the momentum by forming
\begin{eqnarray*}
\rho(x,t)P_M(x,t)=\int pF_\psi(p,x,t)dp
\end{eqnarray*}
which is exactly the Bohm momentum $P_B$, in other words it is identical to the Moyal momentum defined in~\cite{jm49, bh11}.

Let us integrate equation (\ref{eq:mbt}) over $p$ we find
\begin{eqnarray*}
\frac{\partial \rho(x,t)}{\partial t}+\frac{\partial}{\partial x}\left(\rho(x,t)\frac{\partial S(x,t)}{\partial x}\right)=0.
\end{eqnarray*}
This equation is identical to equation (\ref{eq:Liouv}) being the Liouville equation.  If one carries out the same procedure on equation (\ref{eq:bbt}) one finds, after some work, the QHJ equation (\ref{eq:qhj}) including the QPE.  Here projecting into a configuration space is equivalent to finding the marginal properties from a phase distribution $F_\psi(p,x,t)$.  Notice again that the QPE only appears in the projection.   The fact that equation (\ref{eq:bbt}) reduces directly to the classical HJ equation the way it does shows that the QHJ is not an equation that is pulled `out of a hat' as it were but is an essential part of the standard mathematical structure of quantum mechanics.

\section{Conclusion}

We have shown that the Lagrangian of a Schr\"{o}dinger field (\ref{eq:Lagrangian}) can be used to produce two sets of equations of motion via the Euler-Lagrange equations; either the Schr\"{o}dinger equation and its dual or, if we replace $\psi$ by two real fields $R$ and $S$,  the two equations (\ref{eq:Liouv}) and (\ref{eq:qhj}).  Equation (\ref{eq:Liouv}) gives the well known Liouville equation, the other is the quantum Hamilton-Jacobi equation [QHJ] which is rarely referred to in the literature.

 The reason for this could simply be that the Schr\"{o}dinger equation is linear, whereas the QHJ equation is non-linear and therefore difficult to use.  
However this equation has been used by Bohm~\cite{db52} to provide a different, but controversial, interpretation and this may have generated an air of distrust in the equations themselves.   However as far as the mathematics is concerned they are different ways of discussing the {\em same mathematical structure} independent of any specific interpretation.  If used correctly they give exactly the same results as standard quantum mechanics but give a different insight into quantum phenomena.

In this paper we have shown that the energy-momentum tensor, $T^{\mu\nu}$, expressed in terms of real fields provides, a more transparent way of discussing energy flows involved in quantum processes.  Conventional quantum field theory, uses the same energy-momentum tensor but integrates the components over all space before associating them with the energy and momentum of the `particle'.  In other words, conventional QFT discusses the global aspects of the theory with no concern for non-locality, whereas we provide a local description which enables us to bring out more clearly the details of the non-local aspects  of the energy and momentum.  For a recent discussion which highlights some of the same problems see Colosi and Rovelli~\cite{dccr09}.

While global energy conservation is ensured through equation (\ref{eq:conse}), local energy conservation is ensured through the QHJ equation (\ref{eq:qhj}). This means we have a way of keeping track of the energy flow that is closer to, but different from, the way energy flow is expressed in the classical Hamilton-Jacobi equation.  It was this feature that led Bohm~\cite{db52} initially to keep the notion of a localised particle whose progress could be tracked along a trajectory.  But the QHJ equation has many  subtle features that show that his approach is not a return to a classical deterministic mechanics--a point that was extensively discussed by Bohm himself~\cite{db57}.  

For example, there is the appearance of a non-local coupling between particles described by entangled wave functions.  The two-particle equation (\ref{eq:2body}) shows that this coupling arises not because there is a force between the two particles, but because the energy has become delocalised and non-locally associated with both particles, a notion that is anathema in classical physics.  By integrating over space, conventional QFT builds in this non-locality implicitly.  The new feature involving weak values means that it is no longer merely a theoretical choice between local and global values because we are now in a position to determine empirically both sets of values using weak measurements.

The central question discussed in this paper concerns the meaning of the one-parameter solutions of the QHJ equation, solutions that when taken to the classical limit become particle trajectories. In an attempt to provide a local formulation in configuration space we are forced to confront non-locality.  In this paper we traced the reason for this non-locality to the non-commutativity in the theory.   We are led to distinguish between left and right translations, and it is this feature that leads us to a deeper theory based on a non-commutative symplectic (phase) space.  It is when we  project into a configuration sub-space that the QHJ equation as well as the Liouville equation (\ref{eq:Liouv}) appear.  Thus these equations, rather than being fundamental equations in their own right, are fragments of a deeper theory based on a non-commutative phase space. In this theory the particle can no longer be regarded as a point-like entity but is at best a quasi-local structure the nature of which is still under active investigation~\cite{mdg12}\cite{gdmdgbh14}.  A  more technical and detailed account of the algebraic approach can be found in Hiley~\cite{bh11d}.  Until these details have been well understood, any attempt to provide a final interpretation of these equations, and in fact the Schr\"{o}dinger equation itself, are bound to be inadequate.



\bibliography{myfile}{}
\bibliographystyle{plain}

\end{document}